\documentclass[12pt]{iopart}
\usepackage{iopams}
\usepackage{graphicx}

\newcommand{\vth}{\vartheta}
\newcommand{\dst}{\displaystyle}
\newcommand{\tst}{\textstyle}
\newcommand{\rhm}{\rho_{m}}
\newcommand{\prm}{p_{m}}
\newcommand{\dph}{\dot{\phi}}
\newcommand{\ddrhm}{\frac{d\phantom{\rho}}{d\rhm}}
\newcommand{\pat}{\partial}

\newcommand{\const}{\mathrm{const}}

\begin{document}
\title{On gravitational repulsion effect at extreme conditions in gauge theories of gravity}
\author{A. V. Minkevich}
\address{Department of Theoretical Physics, Belarussian State University, F. Skoriny av. 4,
Minsk 220030, Belarus}
\address{ Department of Physics and Computer Methods, Warmia
and Mazury University in Olsztyn, Poland}
\begin{abstract}
Homogeneous isotropic gravitating models are discussed in the framework of gauge approach to
gravitation. Generalized cosmological Friedmann equations without specific solutions are deduced
for models filled by scalar fields and usual gravitating matter. Extreme conditions, by which
gravitational repulsion effect takes place, are analyzed.
\end{abstract}

\section{Introduction}

As it is well known, general relativity theory (GR) does not lead to restrictions for admissible
energy densities in the case of gravitating systems with positive values of energy density and
pressure satisfying energy dominance condition. As a result, the problem of gravitational
singularities takes place in the frame of GR [1]. The appearance of nonphysical state with
divergent energy density limiting world lines in the past or in the future is characteristic
feature of Friedmannian cosmological models (problem of cosmological singularity - PCS) and also of
collapsing systems. From physical point of view, the problem of gravitational singularities is
connected with the fact that gravitational interaction in the case of gravitating systems with
positive energy densities and pressures in the frame of GR as well as Newton's theory of
gravitation has the character of attraction but not repulsion. Note that gravitational interaction
in GR can have the repulsion character in the case of gravitating systems with negative pressure
(for example systems including massive or nonlinear scalar fields). According to accepted opinion,
it is possible cause of acceleration of cosmological expansion at present epoch. However, such
effect does not permit to solve the PCS in the frame of GR [2]: all Friedmannian cosmological
models of flat and open type, and the most part of closed models are singular.

There were many attempts to solve the problem of gravitational singularities and at first of all
the PCS in the frame of GR and other classical theories of gravitation; a number of particular
regular cosmological solutions was obtained (see [3] and references given herein). In connection
with this note that the solution of PCS means not only obtaining regular cosmological solutions,
but also excluding singular solutions; as a result regular behaviour of cosmological solutions has
to be their characteristic feature. The most existent attempts to solve the PCS do not satisfy
indicated conditions. According to wide known opinion, the solution of PCS and generally of the
problem of gravitational singularities of GR has to be connected with quantum gravitational
effects, which must be essential at Planckian conditions, when energy density is comparable with
Planckian one. Previously some regular bouncing cosmological solutions were obtained in the frame
of candidates to quantum gravitation theory --- string theory/M-theory and loop quantum gravity
(see, for example, [4--6]). From physical point of view, these solutions have some difficulties
[7]. In the case of bouncing cosmological solutions built in the frame of  string theory the
condition of energy density positivity for gravitating matter is violated. In the case of loop
quantum cosmology a bounce takes place for microscopic model having a volume comparable with
Planckian one. If one supposes that Universe at compression stage is macroscopic object, one has to
explain the transformation of macro-universe into micro-universe before a bounce, this means one
has to introduce some model inverse to inflation.

As it was shown in a number of papers (see [7, 16] and references given herein), gauge theories of
gravitation (GTG) and at first of all the Poincare GTG, which are natural generalization of GR by
applying the gauge approach to gravitational interaction [8,9], offer an opportunity to solve the
PCS. All homogeneous isotropic cosmological models including inflationary models are regular in
metrics, Hubble parameter, its time derivative by certain restriction on equation of state for
gravitating matter at extreme conditions. Unlike Friedmannian models of GR nonphysical state with
divergent energy density does not appear because of gravitational repulsion effect at extreme
conditions, which takes place in GTG in the case of usual gravitating systems with positive energy
densities satisfying energy dominance condition.

The present paper is devoted to analysis of gravitational repulsion effect at extreme conditions in
the frame of GTG in the case of homogeneous isotropic gravitating systems.  In Section 2 some
important properties of homogeneous isotropic models in the frame of Poincare GTG are briefly
discussed.  In Section 3 generalized cosmological Friedmann equations without specific solutions
for such models filled by usual matter and scalar fields are deduced. In Section 4 extreme
conditions leading to gravitational repulsion effect are analyzed.

\section{Homogeneous isotropic gravitating models in Poincare GTG}

In the framework of gauge approach to gravitation the Poincare GTG is of the greatest interest
[8,7]. Gravitational field variables in Poincare GTG are the tetrad (translational gauge field) and
Lorentz connection (Lorentz gauge field), corresponding field strengths are torsion and curvature
tensors, and physical space-time is 4-dimensional Riemann-Cartan continuum. As sources of
gravitational field in Poincare GTG are covariant generalizations of energy-momentum and spin
tensors. Unlike gauge Yang-Mills fields, for which the Lagrangian is quadratic in the gauge field
strengths, gravitational Lagrangian of Poincare GTG can include also linear in curvature term
(scalar curvature), which is necessary to satisfy the correspondence principle with GR. By using
sufficiently general expression for gravitational Lagrangian, homogeneous isotropic gravitating
models were investigated in the frame of Poincare GTG in a number papers (see for example [7, 10,
16]). Because of high spatial symmetry gravitational equations depend weakly on the structure of
quadratic part of gravitational Lagrangian that permits to obtain conclusions of general character.
Below some important relations describing homogeneous isotropic models in Poincare GTG will be
given.

Spatial symmetries of Riemann-Cartan space are defined by a set of linearly independent Killing's
vectors, with respect to which the Lie derivatives of metric and torsion tensors vanish (see [1]
Chapter 2, [17]). Homogeneous isotropic models possess 6 linearly independent Killing's vectors and
metric tensor in co-moving system of reference  has the form of Robertson-Walker metrics [17]:
\[
\label{rwmetric}
g_{\mu\nu}=diag\left(\;1,\; -\frac{\displaystyle R^2\left(t\right)}%
{\displaystyle 1-k r^2},\;-R^2\left(t\right)r^2, \;-R^2\left(t\right)r^2\sin^2\vth\,\right),
\]
where $R(t)$ is the scale factor, $k=+1,0,-1$ for closed, flat, open models respectively (the light
velocity $c=1$), spatial spherical coordinates are used. Then the torsion tensor
$S^\lambda{}_{\mu\nu}=-S^\lambda{}_{\nu\mu}$   satisfying symmetry conditions can have the
following non-vanishing components [10]: $S^1{}_{10}=S^2{}_{20}=S^3{}_{30}=S(t)$,
$S_{123}=S_{231}=S_{312}=\tilde{S}(t)\frac{R^3r^2}{\sqrt{1-kr^2}}\sin{\theta}$, where $S(t)$ and
$\tilde{S}(t$) are functions of time. The functions $S$ and $\tilde{S}$ have different properties
with respect to transformations of spatial inversions, namely, the function $\tilde{S}(t)$ has
pseudoscalar character. By supposing $\tilde{S}=0$, we obtain for the curvature tensor
$F^{\mu\nu}{}_{\sigma\rho}=-F^{\nu\mu}{}_{\sigma\rho}=-F^{\mu\nu}{}_{\rho\sigma}$ the following
non-vanishing components: $F^{01}{}_{01}=F^{02}{}_{02}=F^{03}{}_{03}\equiv A$ and
$F^{12}{}_{12}=F^{13}{}_{13}=F^{23}{}_{23}\equiv B$ with
\begin{equation}\label{abc}
A=\frac{\left(\dot{R}-2RS\right)^{\cdot}}{R},
 \qquad\qquad
B=\frac{k+\left(\dot{R}-2RS\right)^{2}}{R^2},
\end{equation}
and Bianchi identities in this case are reduced to the only relation
\begin{equation}
\label{bian} \dot{B}+2H\left(B-A\right)+4AS=0,
\end{equation}
where $H=\frac{\dot{R}}{R}$  is the Hubble parameter, and a dot denotes differentiation with
respect to time.

We will use gravitational Lagrangian in the form
\begin{eqnarray}\fl
{\cal L}_{\rm G}=  f_0\, F+F^{\alpha\beta\mu\nu}\left(f_1\:F_{\alpha\beta\mu\nu}+f_2\:
F_{\alpha\mu\beta\nu}+f_3\:F_{\mu\nu\alpha\beta}\right)+ F^{\mu\nu}\left(f_4\:F_{\mu\nu}+f_5\:
F_{\nu\mu}\right)+f_6\:F^2 \nonumber \\
+S^{\alpha\mu\nu}\left(a_1\:S_{\alpha\mu\nu}+a_2\:
S_{\nu\mu\alpha}\right) +a_3\:S^\alpha{}_{\mu\alpha}S_\beta{}^{\mu\beta},\nonumber
\end{eqnarray}
where $F_{\mu\nu}=F^{\alpha}{}_{\mu\alpha\nu}$, $F=F^\mu{}_\mu$,  $f_i$ ($i=1,2,\ldots,6$), $a_k$
($k=1,2,3$) are indefinite parameters, $f_0=(16\pi G)^{-1}$, $G$ is Newton's gravitational
constant. Gravitational equations for homogeneous isotropic gravitating models (with $\tilde{S}=0$)
are reduced to the system of 3 equations [10], which by using (2) can be written as
\begin{equation}
\label{syst1}
\begin{array}{l}
6f_0B-12f\left(A^2-B^2\right)-3a\left(H-S\right)S=\rho,\\
2f_0\left(2    A+B\right)+4f\left(A^2-B^2\right)-a\left(\dot{S}+HS-S^2\right)=-p,\\
f\left(\dot{A}+\dot{B}\right)+\left[f_0+\frac{1}{8}a+4f
\left(A+B\right)\right]S=0.\\
\end{array}
\end{equation}
where $f=f_1+\frac{1}{2}f_2+f_3+f_4+f_5+3f_6$, $a=2a_1+a_2+3a_3$,   $\rho$ is the energy density,
$p$ is the pressure and the average of spin distribution of gravitating matter is supposed to be
equal to zero. The system of equations (3) leads to cosmological equations without high derivatives
if $a=0$ [10] (see below). Then we find from (3) the curvature functions $A$ and $B$ and torsion
$S$ in the following form
\begin{eqnarray}
\label{oldsol}
A=-{\frac{1}{12f_0}}\,{\frac{\rho+3p-\alpha\left(\rho-3p\right)^2/2}%
{1+\alpha\left(\rho-3p\right)}}\, ,\nonumber\\
B={\frac{1}{6f_0}}\, {\frac{\rho+\alpha\left(\rho
-3p\right)^2/4}{1+\alpha\left(\rho-3p\right)}}\, ,\\
S(t)= -\frac{1}{4}\frac{d}{dt} \ln\left|1+\alpha(\rho-3p)\right| \, ,\nonumber
\end{eqnarray}
where indefinite parameter $\displaystyle \alpha=\frac{f}{3f_0\,^2}$ has inverse dimension of
energy density. Note, that in the case $\tilde{S}\neq 0$ Bianchi identities for homogeneous
isotropic models are reduced to two relations and system of gravitational equations of Poincare GTG
includes 4 equations. However, these equations have solution (4) together with $\tilde{S}=0$
always, if $a=0$ [18].

\section{Generalized cosmological Friedmann equations in GTG}

By using expressions (1) of curvature functions for homogeneous isotropic gravitating models and
the solution (4) of gravitational equations of Poincare GTG we obtain the following generalized
cosmological Friedmann equations (GCFE)
\begin{equation}
\label{5}
\displaystyle{\frac{k}{R^2}+\left\{\frac{d}{dt}\ln\left[R\sqrt{\left|1+\alpha\left(\rho-
3p\right)\right|}\right]\right\}^2 }\displaystyle{ =\frac{8\pi G}{3}\;\frac{\rho+
\frac{\alpha}{4}\left(\rho-3p\right)^2}{1+\alpha\left(\rho-3p\right)} \, ,}
\end{equation}
\begin{equation}\fl
\label{6}
\displaystyle{R^{-1}\,\frac{d}{dt}\left[\frac{dR}{dt}+R\frac{d}{dt}\left(\ln\sqrt{\left|1+\alpha\left(\rho
-
3p\right)\right|}\right)\right]} 
\displaystyle{=-\frac{4\pi G}{3}\;\frac{\rho+3p- \frac{\alpha}{2}\left(\rho-3p\right)^2}{
1+\alpha\left(\rho-3p\right)}\, .}
\end{equation}
 The difference of
(5)--(6) from Friedmannian cosmological equations of GR is connected with terms containing the
parameter $\alpha$. These terms arise from quadratic in the curvature tensor part of gravitational
Lagrangian, which unlike metric theories of gravitation does not lead to high derivatives in
cosmological equations.\footnote{As it was shown in [11,12] generalized cosmological Friedmann
equations (5)--(6)can be deduced also in the frame of the most general GTG - affine-metric GTG
[9].} The conservation law in GTG has usual form
\begin{equation}
\label{7} \dot{\rho}+3H\left(\rho+p\right)=0.
\end{equation}

Now by using GCFE (5)-(6) we will consider homogeneous isotropic models
filled by non-interacting
scalar field $\phi$ minimally coupled with gravitation and gravitating
matter with equation of state in
general form $p_m=p_m(\rho_m)$. (The generalization for the case with
several scalar fields can be
made directly). Then the energy density $\rho$ and pressure $p$ take the
form
\begin{equation}
\label{8_} \rho=\frac{1}{2}\dot{\phi}^2+V+\rho_m \quad (\rho>0), \quad
p=\frac{1}{2}\dot{\phi}^2-V+p_m,
\end{equation}
where $V=V(\phi)$ is a scalar field potential. By using the scalar field
equation in
homogeneous isotropic space
\begin{equation}
\label{9} \ddot{\phi}+3H\dph=-\frac{\pat V}{\pat\phi}
\end{equation}
we obtain from  (7)-(9) the conservation law for gravitating matter
\begin{equation}
\label{10} \dot{\rho}_m+3H\left(\rhm+\prm\right)=0.
\end{equation}
By using (8)--(10) the GCFE (5)--(6) can be transformed to the following form
\begin{eqnarray}\fl
\label{11}
 \left\{
 H
\left[
 Z+3\alpha
 \left(
   \dph^2+\frac{1}{2} Y
 \right)
\right]
 +3\alpha\frac{\pat V}{\pat\phi}\dph\right\}^2
 +\frac{k}{R^2}\,Z^2 \nonumber\\
 =\frac{8\pi G}{3}\,
 \left[
   \rhm+\frac{1}{2}\dph^2+V+\frac{1}{4}\alpha\,
   \left(4V-\dph^2+\rhm-3\prm\right)^2
 \right]
\,Z,
\end{eqnarray}
\begin{eqnarray}\fl
\label{12}
\dot{H}\left[
 Z+3\alpha
 \left(
   \dph^2+\frac{1}{2} Y
 \right)
\right]Z  +H^2 \left\{
 \left[
   Z-15\alpha\dph^2
   \right.\right. \nonumber\\
   \left.\left.
   -3\alpha Y-\frac{9\alpha}{2}
        \left(\frac{d\prm}{d\rhm}Y
         +3\left(\rhm+\prm\right)^2 \frac{d^2\prm}{d\rhm^2}
         \right)
 \right]Z
 -18\alpha^2
 \left(\dph^2+\frac{1}{2}Y \right)^2
\!\right\}
 \nonumber\\
-12\alpha H\frac{\pat V}{\pat\phi}\dph
\left[
    Z+ 3\alpha\left(\dph^2+\frac{1}{2}Y\right)
\right]
+3\alpha \left[
    \frac{\pat^2 V}{\pat\phi^2}\dph^2-\left(\frac{\pat V}{\pat\phi}\right)^2
\right]Z \nonumber\\
-18\alpha^2\left(\frac{\pat V}{\pat\phi}\right)^2\dph^2
 =\frac{8\pi G}{3} \left[
    V-\dph^2-\frac{1}{2}\left(\rhm+3\prm\right)
    \right. \nonumber\\
    \left.
    +\frac{1}{4}\alpha\left(4V-\dph^2+\rhm-3\prm\right)^2
\right]Z,
\end{eqnarray}
 where $Z=1+\alpha\left(4V-\dph^2+\rhm-3\prm\right)$ and $Y=\left(\rhm+\prm\right)%
\left(3\frac{d\prm}{d\rhm}-1 \right)$.

By transformation of GCFE (5)-(6) to the form (11)-(12) these equations were multiplied by $Z$. As a
result  (11)-(12) have specific solutions when $Z=0$ [13, 7]. In fact, by using the expression of the
Hubble parameter following from  (11)
\begin{equation}
\label{13} H_{\pm}=\frac{\dst -3\alpha\,\frac{\pat V}{\pat \phi}\, \dot\phi
\pm \sqrt{D}}{ \dst Z+3\alpha\left(\dph^2+\frac{1}{2}Y\right)},
\end{equation}
where
$$
\ D=\frac{8\pi G}{3}\,\left[\rhm+\frac{1}{2}\dph^2+V
 +\frac{1}{4}\alpha\,
\left(4V-\dph^2+\rhm-3\prm\right)^2\right]\,Z-\frac{k}{R^2}Z^2,
$$
it is easy to show that  (12) is satisfied, if $Z=0$. It is because the terms in (12), which don't
include $Z$ as multiplier, vanish by virtue of (13). Excluding these terms and dividing obtained
equation on $Z$ we will have, instead of (12), the following equation without specific solutions
\begin{eqnarray}\fl
\label{14}
\dot{H}\left[
 Z+3\alpha
 \left(
   \dph^2+\frac{1}{2} Y
 \right)
\right] +3H^2
 \left[
   Z-\alpha\dph^2
   +\alpha Y
   \vphantom{\frac{d^2\prm}{d\rhm^2}}
   \right. \nonumber\\
   \left.
   -\frac{3\alpha}{2}
        \left(\frac{d\prm}{d\rhm}Y
         +3\left(\rhm+\prm\right)^2 \frac{d^2\prm}{d\rhm^2}
         \right)
 \right]
+3\alpha \left[
    \frac{\pat^2 V}{\pat\phi^2}\dph^2-\left(\frac{\pat V}{\pat\phi}\right)^2
\right] \nonumber\\
=8\pi G \left[
    V+\frac{1}{2}\left(\rhm-\prm\right)
    +\frac{1}{4}\alpha\left(4V-\dph^2+\rhm-3\prm\right)^2
\right]-\frac{2k}{R^2}Z.
\end{eqnarray}
 The system of  equations (11) and (14) is equivalent to GCFE (5)-(6) in considered case of models filled
by usual matter and scalar fields. If the interaction between scalar fields and usual gravitating
matter is not neglecting, is allowed the GCFE (11), (14) have to be generalized. By taking into
account the interaction by means of scalar field potentials, which in this case depend also on the
energy density of gravitating matter $\rhm$, namely $V=V(\phi, \rho_m)$ [7], we can obtain the
generalization of  equations (11), (14) by similar way. The form of  (11) does not change, but the
value of $Y$ in this case is defined as $Y=\frac{\tst \rhm+\prm}{\tst 1+\frac{\pat V}{\pat \rhm}}
\left( \ddrhm\left( 3\prm-4V \right)-1 \right)$. The equation (11) is
generalized as
\begin{eqnarray}\fl
\label{15}
\dot{H}\left[
 Z+3\alpha
 \left(
   \dph^2+\frac{1}{2} Y
 \right)
\right]  +3H^2
 \left[
   Z-\alpha\dph^2
   \vphantom{\frac{\tst \rhm+\prm}{\tst 1+\frac{\pat V}{\pat \rhm}}}
   +\frac{5\alpha}{2} Y
   \right. \nonumber\\
   \hspace{-4em}\left.
   -\frac{3\alpha}{2}
        \frac{Y}{1+\frac{\pat V}{\pat\rhm}}
        \left(
            1+\frac{d\prm}{d\rhm}+\frac{\rhm+\prm}
                    {1+\frac{\pat V}{\pat\rhm}}
                \frac{\pat^2 V}{\pat\rhm^2}
        \right)
         -\frac{3\alpha}{2}
            \frac{\left(\rhm+\prm\right)^2}{\left(1+\frac{\pat V}{\pat\rhm}\right)^2}
            \frac{d^2}{d\rhm^2}
            \left(3\prm-4V\right)
 \right]
 \nonumber\\
 \hspace{-4em}
- \frac{27}{2}\,\alpha H\dph\,
        \frac{\tst \rhm+\prm}{\left(\tst 1+\frac{\pat V}{\pat \rhm}\right)^2}\,
            \frac{\pat^2 V}{\pat\phi\pat\rhm}
            \left[1+\frac{1}{3}\ddrhm\left(\prm+2V\right)\right]
+3\alpha \left[
    \frac{\pat^2 V}{\pat\phi^2}\dph^2-\left(\frac{\pat V}{\pat\phi}\right)^2
\right]
 \nonumber\\
=8\pi G \left[
    V+\frac{1}{2}\left(\rhm-\prm\right)
    \right.
    \left.
    +\frac{1}{4}\alpha\left(4V-\dph^2+\rhm-3\prm\right)^2
\right]-\frac{2k}{R^2}Z.
\end{eqnarray}
 The system of  (11), (15) does not have specific solutions (with $Z=0$) unlike  (9)-(10) of
 reference [7]. By using obtained equations (11), (14), (15) the repulsion gravitational effect can be analyzed
in the case of homogeneous isotropic gravitating systems in the frame of GTG.

\section{Gravitational repulsion effect in GTG}

As it was noted above, the difference of GCFE from Friedmannian cosmological equations of GR is
connected with terms containing the parameter $\alpha$. The value of $|\alpha|^{-1}$ determines the
scale of extremely high energy densities. Solutions of GCFE (5)-  (6) coincide practically with
corresponding solutions of GR if the energy density is small $\left|\alpha(\rho-3p)\right|\ll 1$
($p\neq\frac{1}{3}\rho$). The difference between GR and GTG can be essential at extremely high
energy densities $\left|\alpha(\rho- 3p)\right|\gtrsim 1$. Ultrarelativistic matter
($p=\frac{1}{3}\rho$) and gravitating vacuum ($p=- \rho$) with constant energy density are two
exceptional systems because  GCFE (5)--(6) are identical to Friedmannian cosmological equations of
GR in these cases independently on values of energy density, and non-einsteinian space-time
characteristics vanish. Properties of solutions of  equations (5)--(6) at extreme conditions depend
on the sign of parameter $\alpha$ and certain restriction on equation of state of gravitating
matter. The study of inflationary models including scalar fields shows that GCFE (5)--(6)lead to
acceptable restriction for scalar field variables if $\alpha>0$ [7, 16]. In the case $\alpha>0$ all
cosmological solutions including inflationary solutions have regular bouncing character if at
extreme conditions $p>\frac{1}{3}\rho$. There are physical reasons to assume that the restriction
$p>\frac{1}{3}\rho$ is valid for gravitating matter at extreme conditions [14]. Really in the case
of perfect gas of fermions at zeroth absolute temperature, the pressure tends to the value $\rho/3$
from below, if density of gas increases. Then we have for nuclear matter at extreme conditions
$p>\frac{1}{3}\rho$ because of strong nucleon interaction [14]. We  will suppose below that the
condition $p>\frac{1}{3}\rho$ is valid for gravitating matter at extreme conditions and, in
particular, at the beginning of cosmological expansion. \footnote {In the case $\alpha<0$ GCFE (5)
-- (6) lead to regular bouncing solutions if at extreme conditions $p<\frac{1}{3}\rho$ [10, 3].
However, in this case GCFE have also singular solutions for some hypothetical superdense systems
[15, 3].} Note, that this condition is valid for so-called stiff equation of state $p=\rho$ used in
the theory of the early Universe (Ya.~B.~Zeldovich and others).

The GCFE lead to restrictions on admissible values of energy density. In fact, if energy density
$\rho$ is positive and $\alpha>0$, from equation (5) in the case $k=+1$, $0$ follows the relation:
\begin{equation}
\label{16} Z\equiv 1+\alpha\left(\rho-3p\right)\ge 0.
\end{equation}
The condition (16) is valid not only for closed and flat models, but also for cosmological models
of open type ($k=-1$) [7]. In the case of models filled by usual gravitating matter without scalar
fields the equation $Z=0$ determines limiting (maximum) energy density, and regular transition from
compression to expansion (bounce) takes place for all cosmological solutions by reaching limiting
energy density. Near limiting energy density the gravitational interaction has the repulsion
character. In the case of systems including also scalar fields a bounce takes place in points of
so-called "bounce surfaces" in space of variables ($\phi$, $\dot\phi$, $\rhm$) [7]. Near bounce
surfaces as well as bounds $Z=0$ gravitational interaction has the character of repulsion, but not
attraction. By using GCFE in the form (11), (14) we will find below, by what conditions
gravitational repulsion effect takes place.

a) At first we will consider gravitating systems filled by gravitating matter without scalar
fields. Then the acceleration $a=\frac{\ddot R}{R}=\dot{H}+H^2$ from  (13)-(14) takes the following
form
\begin{eqnarray}
\fl
a=\left(Z+\frac{3}{2}\alpha Y\right)^{-3}\!
    \left\{\left[4\pi G
            \left(\rho_m-p_m+\frac{\alpha}{2}(\rho_m-3p_m)^2\right)
            -\frac{2k}{R^2}Z
        \right]\left(Z+\frac{3}{2}\alpha Y\right)^2
        \right. \nonumber\\
        \left.
        -2D
        \left[Z+\frac{3}{4}\alpha Y
            -\frac{9\alpha}{4}\left(\frac{d\prm}{d\rhm} Y+3(\prm+\rhm)^2\frac{d^2\prm}{d\rhm^2}\right)
        \right]
    \right\}.
\end{eqnarray}
Obviously the repulsion ($a>0$) will take place, if the expression in figured parentheses in (17)
is positive. The domain of admissible energy densities, by which the repulsion effect takes place,
depends on equation of state of gravitating matter at extreme conditions and on the value of
parameter $\alpha$. In particular, in the case of flat models $(k=0)$ with linear equation of state
$p_m = w\rho_m$ $(w=\const>\frac{1}{3})$ we obtain from (17) the following condition for energy
densities corresponding to repulsion effect
\begin{equation}
y(x)\equiv \frac{1}{8}(9w^2-1)x^3+\frac{3(3w^2-1)}{3w+1}x^2+\frac{3(9w+5)}{2(3w+1)}x -1>0 ,
\end{equation}
where $x=\alpha\rhm(3w-1)>0$. Moreover from condition $Z=1-x\ge 0$ follows, that $x\le 1$. By this
restriction the cubic equation $y(x)=0$ has the only real root $x_0$ and inequality (18) is
fulfilled at $x>x_0$. Numerical solution of equation $y(x)=0$ gives the dependence of value $x_0$
on parameter $w$  (see Fig.1). The gravitational repulsion effect in considered case takes place at
energy densities defined by the following condition $x_0<x\le 1$, and the value $x=1$ corresponds
to limiting energy density.
\begin{figure}[hbt]
\centering{
\includegraphics[width=0.5\textwidth]{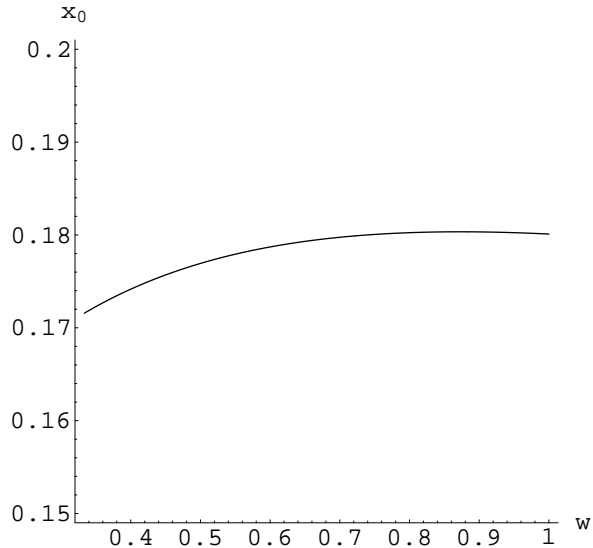}
\caption{\label{mnfig2} The root $x_0$ as function of parameter $w$.} }
\end{figure}

b) In the case of systems including non-interacting gravitating matter and scalar field the
condition of gravitational repulsion obtained from (11) and (14) is
\begin{eqnarray}\fl
\left\{8\pi G
    \left[
        V+\frac{1}{2}\left(\rhm-\prm\right)
        +\frac{\alpha}{4}(4V-\dph^2+\rhm-3\prm)^2
    \right]
    \right. \nonumber\\
    \left.
    -3\alpha
        \left[
            \frac{\pat^2 V}{\pat\phi^2}\dph^2-\left(\frac{\pat V}{\pat\phi}\right)^2
        \right]
        -\frac{2k}{R^2}Z
\right\}
\left[Z+3\alpha\left(\dph^2+\frac{1}{2}Y \right)
\right]^2
 \nonumber\\
+\left(-3\alpha\frac{\pat V}{\pat \phi}\dot{\phi}\pm\sqrt{D}\right)^2
    \left[-2Z+6\alpha\dph^2-\frac{3}{2}\alpha Y
            \right. \nonumber\\
            \left.
            +\frac{9\alpha}{2}\left(\frac{d\prm}{d\rhm}Y+3(\prm+\rhm)^2\frac{d^2\prm}{d\rhm^2}\right)
    \right]>0
\end{eqnarray}
Inequality (19) together with condition $Z\ge 0$ determine the domain of variables $\phi$,
$\dot\phi$ and $\rhm$ at extreme conditions near bounds ($Z=0$) and bounce surfaces in space of
these variables, where gravitational repulsion effect appears. This domain is different for $H_+$-
and $H_-$-solutions of (11) and (14) corresponding to two values of the Hubble parameter (13) [7].
Note that  (19) describes also repulsion effect at small energy densities like to GR, when the
pressure is negative because of contribution of scalar field. In the case of systems filled by
interacting gravitating matter and scalar field the generalization of condition (19) can be
obtained from (11) and (15). Note, that in the case of closed and open models conditions for
gravitational repulsion effect include the term with scale factor $R$ depending on 3-space
curvature also, this means that the repulsion effect for homogeneous isotropic systems depends on
global structure of gravitating model.

\section{Conclusion}

The analysis of gravitational repulsion effect at extreme conditions in GTG presented above shows
that this effect takes place near limiting energy density, near a bounce, and it depends on content
and properties of gravitating matter at extreme conditions (equation of state for gravitating
matter, the form of scalar field potentials etc) and also on the scale of extremely high energy
densities defined by parameter $\alpha$. If the value of limiting energy density is essentially
less than the Planckian one, gravitational repulsion effect appears at classical conditions, when
 the quantum gravitational corrections according to widely known opinion are not
essential.\footnote{Probably, in the frame of considered GTG the quantum properties of gravity can
be important near the bound $Z=0$, where the values of the torsion function $S$ are extremely
large.} In the case of systems including scalar fields the value of limiting energy density is
different for different solutions and can be essentially greater than $\alpha^{-1}$, but the
appearance of gravitational repulsion effect does not depend on this fact [16]. If limiting energy
density is comparable with the Planckian one, quantum gravitational corrections have to be taken
into account, although classical GTG ensures satisfactory non-singular behaviour of gravitating
systems. Because GCFE in the case of gravitating systems with small energy densities lead to
consequences similar to that of GR, gravitational repulsion effect at such conditions can appear in
the case of gravitating models with negative pressure (dark energy) like to GR. In particular, if
GCFE (5)--(6) include cosmological constant, the value of which corresponds to dark energy density
at present epoch, regular cosmological solutions of (5)--(6) will contain the stage of accelerating
cosmological expansion by dominating of vacuum energy (cosmological constant).

The study of gravitating systems in the frame of GTG shows, that important features of
gravitational interaction at extreme conditions depend essentially on properties of gravitating
matter, which are determined by other fundamental physical interactions. To describe the evolution
of gravitating models we have to know how corresponding properties of gravitating matter (at first
of all its equation of state) change by model evolution. This conclusion obtained by investigation
of homogeneous isotropic models has sufficiently general character, because the form of used GCFE
(5)-(6) does not depend on detailed structure of quadratic in the curvature part of gravitational
Lagrangian of GTG. At the same time the dynamics of gravitating systems depends on the structure of
quadratic part of gravitational Lagrangian of GTG, if the homogeneity and isotropy are broken. In
connection with this, the search of GTG leading to the most satisfactory consequences in general
case of inhomogeneous anisotropic models is of direct physical interest.

\begin{ack}
The author is very grateful to Dr. A. S. Garkun and Dr. A. A. Minkevich for the help in preparing
of this paper.
\end{ack}


\begin{thebibliography}{99}
\bibitem{ml1} Hawking S W and Ellis G F R 1973 {\it The Large Scale Structure of Space-Time\/}
(Cambridge: Cambridge University Press)
\bibitem{ml2} Starobinsky A A 1978 {\it Pis'ma v Astron. Zhurn.\/}
{\bf 4} 155; Belinsky V A, Grishchuk L P, Zeldovich Ya B, Khalatnikov I M 1985  {\it Zhurn. Exp.
Teor. Fiz.\/} {\bf 89} 346
\bibitem{ml3} Minkevich A V 1998 {\it Acta Phys. Polon.\/} B {\bf 29} 949.
\bibitem{ml4} Gasperini M, Veneziano G 2003 {\it Phys. Rep.\/} {\bf 373} 1 ({\it Preprint\/}
hep-th/0207130)
\bibitem{ml5} Bozza V, Veneziano G 2005 Scalar perturbations in regular two-component bouncing cosmologies {\it
Preprint\/} hep-th/0502047; 2005 Regular two-component bouncing cosmologies and perturbations
therein {\it Preprint\/} gr-qc/0506040
\bibitem{ml6} Bojowald M 2002 {\it Class. Quant. Grav.\/} {\bf 19} 2717 ({\it Preprint\/}
gr-qc/0202077)
\bibitem{ml7} Minkevich A V 2006 {\it Gravitation\&Cosmology} {\bf 12}  no.~1(45) 11 ; ({\it
Preprint\/} gr-qc/0506140);2005 {\it Int. J. Mod.
Phys.\/} A {\bf 20} 2436
\bibitem{ml8} Hehl F W 1980 Four Lectures on Poincare Gauge Field Theories, in: {\it Cosmology and Gravitation\/}
(New York: Plenum Press)
\bibitem{ml9} Hehl F W, McCrea J D, Mielke E W,  and Ne'eman Y 1995 {\it Phys.
Rep.\/} {\bf 258} 1
\bibitem{ml10} Minkevich A V 1980 {\it Vestsi Akad. Nauk BSSR. Ser. fiz.-mat.\/}
no.~2 87; 1980 {\it Phys. Lett.\/} A {\bf 80} 232
\bibitem{ml11} Minkevich A V 1993 {\it Dokl. Akad. Nauk Belarus.\/} {\bf 37} 33
\bibitem{ml12} Minkevich A V and Garkun A S 2000 {\it Class. Quantum Grav.\/} {\bf 17} 3045
\bibitem{ml13} Minkevich A V 2004 {\it Annals of European Acad. Sci.\/}
\bibitem{ml14} Zeldovich Ya B, Novikov I D  1975 {\it Structure and
evolution of the Universe\/} (Moscow: Nauka) 37
\bibitem{ml16} Minkevich A V 1986 {\it Dokl. Akad. Nauk Belarus.}
{\bf 30} 311
\bibitem{ml17} Minkevich A V and Garkun A S 2005 Analysis of inflationary cosmological models in gauge
theories of gravitation {\it Preprint\/} gr-qc/0512130; to be published in {\it Class. Quantum
Grav.} (2006)
\bibitem{ml18} Weinberg S 1972 {\it Gravitation and Cosmology\/} (New York, London, Sydney,
Toronto: John Wiley and Sons, Inc.) Chapter 13
\bibitem{ml19} Minkevich A V 1985 {\it Physical aspects of gauge approach to gravitation theory\/}
(Minsk: BSU)
\end{thebibliography}
\end{document}